# Integer spin-chain antiferromagnetism of the 4*d* oxide $CaRuO_3$ with post-perovskite structure

Y. Shirako,[1,*] H. Satsukawa,[2] X. X. Wang,[3,4] J. J. Li,[3,4] Y. F. Guo,[5] M. Arai,[6] K. Yamaura,[3,4,7]
M. Yoshida,[1] H. Kojitani,[1] T. Katsumata,[1] Y. Inaguma,[1] K. Hiraki,[2] T. Takahashi,[2] M. Akaogi [1]

[1] Department of Chemistry, Gakushuin University, 1-5-1 Mejiro, Toshima-ku, Tokyo 171-8588, Japan
[2] Department of Physics, Gakushuin University, 1-5-1 Mejiro, Toshima-ku, Tokyo 171-8588, Japan
[3] Superconducting Materials Center, National Institute for Materials Science, 1-1 Namiki, Tsukuba, Ibaraki 305-0044, Japan
[4] Department of Chemistry, Graduate School of Science, Hokkaido University, Sapporo, Hokkaido 060-0810, Japan
[5] International Center for Materials Nanoarchitectonics (MANA), National Institute for Materials Science, 1-1 Namiki, Tsukuba, Ibaraki 305-0044, Japan305-0044, Japan
[6] Computational Materials Science Center, National Institute for Materials Science, 1-1 Namiki, Tsukuba, Ibaraki 305-0044, Japan
[7] JST, Transformative Research-Project on Iron Pnictides (TRIP), 1-1 Namiki, Tsukuba, Ibaraki 305-0044, Japan

**Abstract**

A quasi-one dimensional magnetism was discovered in the post-perovskite $CaRuO_3$ ($Ru^{4+}$: $4d^4$, *Cmcm*), which is iso-compositional with the perovskite $CaRuO_3$ (*Pbnm*). An antiferromagnetic spin-chain function with $-J/k_B$ = 350 K well reproduces the experimental curve of the magnetic susceptibility vs. temperature, suggesting long-range antiferromagnetic correlations. The anisotropic magnetism is probably owing to the $d_{yz}$ - $2p_\pi$ - $d_{zx}$ and $d_{zx}$ - $2p_\pi$ - $d_{yz}$ superexchange bonds along *a*-axis. The Sommerfeld coefficient of the specific heat is fairly small, 0.16(2) mJ $mol^{-1}$ $K^{-2}$, indicating that the magnetism reflects localized nature of the 4*d* electrons. As far as we know, this is the first observation of an integer ($S$ = 1) spin-chain antiferromagnetism in the 4*d* electron system.

## I. Introduction

A high-pressure dense phase of $CaRuO_3$ was recently created by high-pressure and high-temperature experiments, and it was successfully quenched to ambient conditions [1,2]. The crystal structure at ambient conditions was identified to be of the post-perovskite type, which is entirely different from the perovskite structure of the regularly synthesized $CaRuO_3$ [1]. The newly synthesized phase is 1.7 % denser in volume than the perovskite $CaRuO_3$ [1]. A quick look of the post-perovskite structure (Fig. 1) indicates that it comprises a $Ca^{2+}$ layer and $RuO_3^{2-}$ layer; which alternatively stacks up along *b*-axis (*Cmcm* [1]). The $RuO_3$ layer consists of $RuO_6$ octahedra ($Ru^{4+}$: $4d^4$) connecting to neighbors by sharing common edges and corners along *a*- and *c*-axis, respectively. The highly anisotropic structure suggests that a much anisotropic electronic state compared with the known perovskite $CaRuO_3$. We thus decided to study the possible anisotropic properties because, generally, new and valuable correlated properties such as high-$T_c$ superconductivity induced by metallization via chemical doping or physical squeezing can be expected for such the anisotropically correlated materials.

Early studies concluded that $CaRuO_3$ was a paramagnetic metal [4-6]. In 1990s, however, NMR studies and development of theoretical understanding of itinerant electrons revealed that it is rather a nearly ferromagnetic metal than the paramagnetic metal, although it does not show a magnetically long-range order [7-10]. In addition, disorder induced ferromagnetic behavior was observed in 2001, suggesting that $CaRuO_3$ is indeed in the vicinity of a magnetic critical point between ferromagnetic and paramagnetic states [11,12]. However, a magnetic glassy was found for a high-quality crystal by Mössbauer spectroscopy [13]. Besides, a possible hidden magnetic order was suggested by anomalous Hall-effect [14]. It appears that the magnetically critical behavior of the perovskite $CaRuO_3$ is still highly under debated [15-21].

To solve the alleged contradictions, it could be helpful not only to study the perovskite family $ARuO_3$ ($A$ = Ca, Sr, Ba [22]) but also to investigate related 4*d* oxides such as $Sr_3Ru_2O_7$ [23], $La_4Ru_6O_{19}$ [24], $Sr_2RuO_4$ [25,26], $SrRhO_3$ [27], and $CaRhO_3$ [28,29], which also demonstrate interesting electronic features. We think a magnetic study of the post-perovskite $CaRuO_3$ can also help to address the issues regarding the magnetically critical behavior of the perovskite $CaRuO_3$. Zhong *et al* [30,31] have studied ground state properties of the post-perovskite $CaRuO_3$ from first principles calculations within generalized gradient approximation (GGA) and GGA+U, which incorporates additional on-site Coulomb interaction between 4*d* orbitals on Ru atoms. From these studies, they found that metallic ferromagnetic state is stable within GGA and insulating AFM state becomes stable for GGA+U with $U \geq 2$ eV.

In this study, we found a quasi-one dimensional (1D) magnetism that is indicative of long-range antiferromagnetic (AFM) correlations in the post-perovskite $CaRuO_3$. As a result, this is the first observation of an integer ($S = 1$) AFM spin-chain among 4*d* correlated oxides.

## II. Experimental

A polycrystalline dense pellet (the observed density indicated <3 % porosities) of the post-perovskite $CaRuO_3$ was synthesized using a Kawai-type multi-anvil high-pressure apparatus at Gakushuin University [1]. First, we synthesized a precursor; mixed powder of $CaCO_3$ and $RuO_2$ was heated at 1150 ºC for 14 hours in air using a Pt crucible. The product was confirmed to be of the perovskite $CaRuO_3$ by powder X-ray diffraction (XRD). Afterward, the powder was sealed in a Pt tube (1.5 mm in diameter and 5.7 mm in length). The tube was placed in the high-pressure apparatus with an MgO pressure medium and a $LaCrO_3$ thermal insulator; details of the high-pressure cell were reported elsewhere [1]. Tungsten carbide anvils with truncated edge length of 2.5 mm were used. Temperature was measured by a Pt-Pt/13 % Rh thermocouple. The synthesis temperature was between 1000 ºC and 1040 ºC, and the synthesis pressure was held constant at 23.5 GPa for a range between 1 and 5 hours. The sample was quenched to room temperature within a few seconds before releasing the pressure. The recovered sample was confirmed to be of the post-perovskite $CaRuO_3$ by XRD rather than the perovskite $CaRuO_3$ [1].

Electrical resistivity ($\rho$) of the post-perovskite $CaRuO_3$ was measured by a four-probes technique with a gauge current of 5 μA on cooling and 50 nA on warming. Electrical contacts on the four locations were made on the pellet by silver wires and paste. Magnetic susceptibility ($\chi$) was measured in a fixed magnetic field of 1, 10, and 50 kOe in a commercial magnetometer (MPMS, Quantum Design). The polycrystalline sample was cooled to 2 K without applying the magnetic field and then warmed to 300 K in the field (zero-field cooling, ZFC), followed by cooling (field cooling, FC). The measurements were repeated in an oven in MPMS between 300 K and 600 K as well. Isothermal magnetization was measured in MPMS at 10 K, 100 K, 200 K, 300 K, and 600 K between -70 kOe and 70 kOe. The heat capacity at constant pressure ($C_p$) was measured in a commercial apparatus (PPMS, Quantum Design) by a quasi-adiabatic method using the same pellet (13.61 mg) after the magnetic and electrical measurements were completed.

After the measurements above, the pellet was ground thoroughly and mixed with an amount of Si powder at approximate volume ratio of 1 to 1. The mixture was spread out evenly on a grease (Apiezon N) coated glass plate. The glass was mounted on a cold stage of a commercial XRD instrument (M03XHF22, Mac Science). Monochromatic radiation (Cu-Kα) was used in a $2\theta$ range between 16 and 96°. The sample temperature was kept constant during the measurements ($|\Delta T| < 1$ K). At every temperature points, the sample position was properly adjusted to avoid tiny mechanical shift due to thermal effect of the cold stage. In addition, tabulated data of the thermal expansivity of Si was used to correct a $2\theta$ shift [32].

## III. Results and Discussion

Temperature effect on $\rho$ of the post-perovskite $CaRuO_3$ is shown in Fig. 2, which clearly indicates that the compound is semiconducting-like over the whole temperature range. At low temperature, $\rho$ goes beyond the technical limit. Both the curves cooling and warming follow exactly the same trace; indicating thermal hysteresis is not obvious over the temperature range. In order to further characterize the $\rho$ feature, we analyzed the data by using the Arrhenius equation; $\rho(T) = \rho_0\exp(E_a/k_BT)$, where $E_a$, $\rho_0$ and $k_B$ are an activation energy, a constant, and the Boltzmann constant, respectively. The theoretical curve fits well the data as shown in the inset to Fig. 2 (solid line corresponds the fitting), estimating $E_a$ to be 0.18 eV.

Fig. 3 shows temperature dependence of $C_p$ of the post-perovskite $CaRuO_3$. It is remarkable that a peak appears at 270 K, which is suggestive of a phase transition. To estimate the transition entropy, $\Delta S$, we first applied a polynomial function by a least-squares-linear method to the $C_p$ data between 220 K and 295 K except the transition part (see the solid curve in the main panel of Fig. 3). Then, we estimated $\Delta S$ to be 0.17 J mol$^{-1}$ K$^{-1}$, as shown in the insets to Fig. 3. The $\Delta S$ is much smaller than the expected $\Delta S$ for full order of the $Ru^{4+}$ spins [$S = 1$; $R$ln3 (~9.13 J mol$^{-1}$ K$^{-1}$), $R$ is the ideal gas constant]; the observed $\Delta S$ corresponds only 1.8 % of the expectation. This suggests that the much of the magnetic entropy is removed via short-range ordering (SRO) above the magnetic transition temperature, if the peak is truly magnetic. Otherwise, a structure change on cooling is somewhat responsible for the observed $\Delta S$. To test the possibility, we carefully conducted low-temperature XRD over the transition temperature.

Fig. 4a shows all the XRD patterns in the order of the temperature. The 300 K pattern was well characterized by the structure model (*Cmcm*) as was done in Ref. 1. We obtained the lattice parameters $a_0$ = 3.114(1) Å, $b_0$ = 9.826(1) Å, $c_0$ = 7.297(1) Å, and $V_0$ = 223.3(1) Å$^3$ by least-squares-linear fitting to the peaks, being identical with the reported values using CrKα radiation [1]. Upon cooling, evident changes such as appearance/disappearance of new peaks and splitting of peaks were not detected, suggesting no symmetry change on the lattice. Fig. 4b shows the temperature dependence of the orthorhombic lattice parameters deduced from the XRD analysis. The lattice parameters anisotropically change on cooling: $b$ and $c$ gradually decrease, while $a$ slightly increases. The anisotropic change is obvious; however the unit-cell volume monotonically decreases without any anomalies. The decrease in volume with temperature is comparable with that of the perovskite $CaRuO_3$ (dotted curve in Fig. 4b [33]). Over the XRD measurements, we found that a structure change is unlikely coupled with the $C_p$ peak at 270 K.

In the low temperature limit of $C_p/T$ vs. $T^2$ (upper inset to Fig. 3), we applied the approximated Debye function $C_v/T = \gamma + \beta T^2$ ($T \ll \Theta_D$) to analyze the part, where $C_v$, $\gamma$, $\beta$ and $\Theta_D$ are heat capacity at constant volume, the Sommerfeld coefficient, a coefficient and the Debye temperature, respectively. By the fitting, we estimated $\gamma$ to be 0.16(2) mJ mol$^{-1}$ K$^{-2}$ and $\beta$ = 4.34(2)×10$^{-5}$ J mol$^{-1}$ K$^{-4}$. $\Theta_D$ was calculated to be 607(3) K from $\beta$. The value of $\gamma$ is fairly smaller than that of the perovskite $CaRuO_3$ (73 mJ mol$^{-1}$ K$^{-2}$ [19]), being qualitatively consistent with the observed semiconducting nature; Conducting carriers in the post-perovskite $CaRuO_3$ are too few to establish a metallic conduction. The observed $\Theta_D$ is slightly higher than the value 495 K found for the perovskite $CaRuO_3$ [8]. It should be noted that a comparable increase in the Debye temperature by the post-perovskite transition was found for $CaRhO_3$ [29], while it was not for $MgSiO_3$ [34].

Fig. 5 shows temperature dependence of $\chi$ of the post-perovskite $CaRuO_3$ measured in a magnetic field of 50 kOe. The curve shows a wide hump with the maximum at 470 K, suggesting a quasi-1D magnetism [35]. In addition, a small kink was found at ~270 K (see the differential curve shown at the bottom of Fig. 5), which is presumably coupled with the $C_p$ peak. To analyze the possible quasi-1D magnetism, we applied the following equation to the experimental curve, which is valid for a 1D AFM Heisenberg spin chain with $S$ = 1 [36];

$$\chi_{\text{calc}} = \frac{N_A \mu_B{}^2 g^2}{k_B T}\left[\frac{2 + 0.0194X + 0.777X^2}{3 + 4.346X + 3.232X^2 + 5.834X^3}\right] + \chi_0 \quad \text{and} \quad X \equiv \frac{|J|}{k_B T} \quad ,$$

where, $N_A$, $\mu_B$, $g$, $J$, and $\chi_0$ are Avogadro number, the Bohr magneton, Landé's general factor, the exchange energy, and a $T$-independent term, respectively. At $\chi_0 = 0$, the agreement factor, $[\sum(\chi_{\text{obs}} - \chi_{\text{calc}})^2/\sum\chi_{\text{obs}}{}^2]^{1/2}$, decreases to 0.00103 over the temperature range from 400 K to 600 K, indicating a good fit. In the analysis, the magnetic parameters $-J/k_B$ and $g$ were estimated to be 350 K ($|J|$ = 30.2 meV) and 2.053, respectively. At $\chi_0 \neq 0$, $-J/k_B$, $\chi_0$, and the agreement factor were estimated to be 350 K, $3.99\times10^{-5}$ emu/mol, and 0.00098, respectively ($g$ was fixed at 2), indicating there is no significant difference between the magnetic parameters with and without $\chi_0$.

The excellent agreement between the experimental and theoretical curves implies that the 1D AFM magnetism is likely above 270 K. It is possible that 1D spin chain is formed along $a$-axis via the Ru-O-Ru bond that has significant $\pi$-type inter-orbital interactions; we will discuss this later. However, disagreement between the theoretical and experimental curves in $T$ vs. $\chi$ is obvious below approximately 350 K. Considering the absence of structure anomaly and $\rho$-$T$ anomaly in the vicinity of 270 K, the disagreement is likely due to establishment of an AFM order in long-range. If this is true, $T_N$ is 270 K, which is 3 times higher than $T_N$ of the post-perovskite $CaRhO_3$ [29].

In Fig. 6, temperature dependence of $\chi$ of the post-perovskite $CaRuO_3$ measured in a weaker magnetic field of 1 kOe is shown. The magnetic anomaly at ~270 K is much obvious. In addition, another anomaly at ~90 K appears with remarkable thermal-hysteresis between the FC and ZFC curves. Further, an additional small peak can be seen at ~4 K, which is confirmed in a 10 kOe measurement (inset to Fig. 6). These multiple anomalies appear only in weaker fields and those are reminiscent of what was observed for the perovskite $CaRuO_3$, for which a spin-glass-like order was suggested [13]. Uni-axial strain in the crystal possibly accounts for the magnetically glassy behaviors, as discussed in Refs. 37 and 38. If this is true, we may expect a similar mechanism for the post-perovskite $CaRuO_3$. Alternatively, it is possible that undetected magnetic impurities are responsible for the anomalies to some extent, such as a possible inter-mediate phase between the perovskite and the post-perovskite as recently found in the Ca-Rh-O system [39].

To characterize the magnetic properties further, we measured isothermal magnetization at various temperatures in a wide temperature range; the data are presented in Fig. 7. We observed only linear features regardless of the magnetic transition at 270 K. This directly indicates that the post-perovskite $CaRuO_3$ does not include ferromagnetic interactions in the magnetic system unlike the perovskite $CaRuO_3$. It is thus interesting to search for a ferromagnetic interaction possibly hidden in the post-perovskite $CaRuO_3$ by introducing disorders as was done for the perovskite $CaRuO_3$ [11,12]; the study is in progress.

Here we compare the magnetic properties of the post-perovskite $CaRuO_3$ ($Ru^{4+}$: $4d^4$) with those of the analogous oxides $CaIrO_3$ ($Ir^{4+}$: $5d^5$) [40,41], $NaIrO_3$ ($Ir^{5+}$: $5d^4$) [42], and $CaRhO_3$ ($Rh^{4+}$:

$4d^5$) [28,29]. Numerical comparison is provided in Table I. We actually investigated 5 post-perovskite oxides synthesized so-far $CaRuO_3$, $CaRhO_3$, $NaIrO_3$, $CaIrO_3$, and $CaPtO_3$; however $CaPtO_3$ was confirmed not to be magnetically active, as was expected ($Pt^{4+}$: $5d^6$; $t_{2g}{}^6$, $S = 0$) [43,44]. Thus, we excluded $CaPtO_3$ from the comparison.

Unlike $CaRuO_3$, the post-perovskite $CaIrO_3$ and $CaRhO_3$ show Curie-Weiss (CW) paramagnetism. Because each compound has a well-localized electronic state, the CW parameters (the Weiss temperature, $\Theta_W$ and the effective magnetic moment, $\mu_{eff}$) can be good indicators of the magnetic state; the CW parameters suggest that AFM interactions play a dominant role between spins. The AFM interactions reasonably account for the AFM long-range order observed for $CaIrO_3$ ($T_N$ = 115 K) and $CaRhO_3$ (90 K). Meanwhile, a relatively small degree of the transition entropy at $T_N$ (4.4 % of the full spin order $\Delta S$ for $CaRhO_3$) suggests that an AFM SRO is significant above $T_N$ [29]. $NaIrO_3$ was also reported to show CW features without long range AFM order (no evidence of magnetic order to 2 K), however it was argued that small amount of magnetic impurities may be responsible for the feature [42]. As a result of the comparison, it becomes clear that the post-perovskite $CaRuO_3$ shows a distinct magnetism.

The microscopic origin of the quasi-1D magnetism of the post-perovskite $CaRuO_3$ may better be discussed after additional studies by microscopic probes such as NMR and μSR are completed; nevertheless considering of the electronic configuration and the structure can reasonably provide a microscopic picture of the quasi-1D magnetism. Looking at the Ru - O bond in the $RuO_6$ octahedra, 4 bonds are 2.039(2) Å and 2 bonds are 1.947(2) Å, indicating a 4.7 % in difference [1]. Thus, it can be expected that the axial compression along $z$ direction which is defined as parallel to the Ru-O1 bond in each $RuO_6$ octahedra [1] should render the $4d_{xy}$ level lower in energy than the $4d_{yz}$ and $4d_{zx}$ level (the $4d_{yz}$ and $4d_{zx}$ orbitals are nearly degenerate). Therefore, it is reasonably expected that the magnetic ground state is $S = 1$ ($4d_{xy}{}^2$ $4d_{yz}{}^1$ $4d_{zx}{}^1$) for the post-perovskite $CaRuO_3$.

To verify these qualitative arguments, we examined electronic structures of the post-perovskite $CaRuO_3$ from first-principles calculations within density functional theory. The full-potential APW method was employed with WIEN2k package [45]. The exchange correlation energy is treated by GGA with the Perdew-Burke-Ernzerhof functional [46]. In Fig. 8a, we show total and partial densities of states (DOS) for non spin-polarized calculations. From the partial DOS (PDOS) decomposed to individual $d$ orbitals on Ru atom, we clearly see that Ru $4d$ bands split into the empty $e_g$ ($3z^2$-1, $x^2$-$y^2$) and the partially filled $t_{2g}$ ($xy$, $yz$, $zx$) bands. In addition, the PDOS of $d_{xy}$ distributes at lower energy than those of $d_{yz}$ and $d_{zx}$. Such splitting is caused by the aforementioned distortion of the $RuO_6$ octahedra. Hence, the Ru 4$d$ configuration can be described as $S$ = 1 ($4d_{xy}^2\ 4d_{yz}^1\ 4d_{zx}^1$).

When magnetic orderings are included in first-principles calculations, the state with magnetic moments aligned antiferromagnetically along $a$-aixs and $c$-axis was found to be more stable than the non spin-polarized and the ferromagnetic states by 8 meV and 4 meV per Ru atom, respectively. For this AFM alignment, we must construct a super cell which is doubled along $a$-axis. This state has not been taken into account in the previous studies, while several other magnetic orderings have been discussed [30,31]. From the PDOS shown in Fig. 8b, we found that the magnetic moments arise from the exchange splittings of $d_{yz}$ and $d_{zx}$ levels and both majority and minority spin bands of $d_{xy}$ are occupied. Thus, the distribution of PDOS indicates that Ru 4$d$ is indeed in the low spin $S$ = 1 configuration. We should note, however, that the calculated AFM states fail to reproduce the experimentally observed insulating behavior. The system is calculated as metallic even though the Fermi energy located in a pseudo gap caused by magnetic ordering. Such disagreement is well known as a failure of GGA. When on-site Coulomb interaction was included as additional terms in GGA+U, the AFM phase became insulating for $U$ ~ 1 eV without changing the splitting between $d$ levels. Detailed results with GGA+U will be reported elsewhere.

The studies above suggest that the magnetic infinite chain is formed along $a$-axis by sharing the common edges of $RuO_6$ octahedra. The nearly equivalent superexchange paths $d_{yz}$ - $2p_\pi$ - $d_{zx}$ and $d_{zx}$ - $2p_\pi$ - $d_{yz}$ connect the neighbor $RuO_6$ octahedra magnetically more strongly along the $a$-axis than the other directions, resulting in the magnetic anisotropy toward 1D. The magnetic bond along the $c$-axis is mainly responsible for the Ru-O-Ru bond with 139°, which is through the corner shared $RuO_6$ octahedra [1]. The neighbor Ru-Ru distance is 18% longer along the $c$-axis than that along the $a$-axis [1]. Thus, the $c$-axis magnetic path can be expected less stronger, rejecting possibility of an in-plane 2D magnetism. Meanwhile, it is reasonable to expect that the $t_{2g}^5$ compounds such as $CaRhO_3$ and $CaIrO_3$ do not form a comparable magnetic path because one of the $d_{xz}$ or $d_{yz}$ orbitals is filled by paired 4$d$-electrons. The $d_{xz}$ and $d_{yz}$ are almost energetically degenerate; thus freedom kills any long-range AFM correlation [47,48]. This fundamentally accounts for absence of the quasi-1D magnetism in the $t_{2g}^5$ compounds $CaIrO_3$ and $CaRhO_3$.

We should state here our general picture for the magnetic interactions of the post-perovskite compounds: 1) quasi-1D AFM correlations are formed along the edge-shared octahedral direction ($a$-axis), 2) freedom of the twofold-degenerate $d_{xz}$ and $d_{yz}$ orbitals plays the predominant role of establishment of the long-rang magnetic order, and 3) minor AFM interactions work along the corner-shared octahedral direction (*c*-axis).

In conclusion, we found the post-perovskite $CaRuO_3$ shows a quasi-1D magnetism owing to the superexchange magnetic paths $d_{yz}$ - $2p_{\pi}$ - $d_{yz}$ and $d_{zx}$ - $2p_{\pi}$ - $d_{zx}$ along the *a*-axis. As a result, this is the first example of integer spin-chain magnetism for a 4*d* electron system. A half-integer ($S$ = 1/2) spin chain was recently discovered for the 4*d* compound $AgSO_4$ [49], in which extended superexchange interactions [50] via ($Ag^{2+}$ - O) - (O - $Ag^{2+}$) with $-J/k_B$ = 217 K play a dominant role to form a Heisenberg AFM chain. Because the magnetic ground states of integer and half-integer AFM spin chains are entirely different [51,52,53], further studies of both the 4*d* compounds can help to understand the underlying physics of 4*d* spin-chain.

The AFM correlation of $CaRuO_3$ is much stronger ($-J/k_B$ = 350 K) than that of $AgSO_4$. Thus, it is of great interest to investigate new magnetic properties those may arise from carrier doping of this novel 4*d* system. We are actively pursuing high pressure experiments to establish the reaction conditions that will result in doping. Besides, the post-perovskite $CaRuO_3$ is unique in that it is iso-compositional but not isostructural with the perovskite $CaRuO_3$. Further study of the post-perovskite $CaRuO_3$ can help to define the fundamental nature of magnetism in the ruthenate family.

**Acknowledgments**

We thank Drs. K. Ohgushi (U. Tokyo), Y. Tsujimoto (NIMS), and J. S. Zhou (U. Texas) for discussion. This research was supported in part by “WPI Initiative on Materials Nanoarchitectonics” from the Ministry of Education, Culture, Sports, Science and Technology, Japan; “Research Seeds Quest Program,” managed by the Japan Science and Technology Agency; and “Grants-in-Aid for Scientific Research (20360012, 22246083, 22340163, 21360325, 20110002, 21540497)” and “Funding Program for World-Leading Innovative R&D on Science and Technology (FIRST Program),” managed by the Japan Society for the Promotion of Science.

## References

* Corresponding author; e-mail: 09242001@gakushuin.ac.jp (Y. Shirako)

Table I. Comparison of the magnetic parameters of the post-perovskite $CaRuO_3$ with those of the related post-perovskite and perovskite oxides.

| Compounds [a] | | | $S$ | $\Theta_W$ (K) | $\mu_{eff}$ ($\mu_B$) | $M_0$ ($\mu_B$) [b] | $T_N$ (K) | Magnetic feature [c] | $E_a$ (eV) [d] | Ref. |
|---|---|---|---|---|---|---|---|---|---|---|
| PPv | $CaRuO_3$ | $Ru^{4+}$: $4d^4$ ($t_{2g}^4$) | 1 | - | - | - | 270 | AFM | 0.18 | This work |
| | $CaRhO_3$ | $Rh^{4+}$: $4d^5$ ($t_{2g}^5$) | 1/2 | -1071 | 2.99 | 0.03 | 90 | CAFM | 0.038 | [29] |
| | $NaIrO_3$ | $Ir^{5+}$: $5d^4$ ($t_{2g}^4$) | 1 | -2.2 | 0.28 | - | - | CWP | VRH | [42] |
| | $CaIrO_3$ | $Ir^{4+}$: $5d^5$ ($t_{2g}^5$) | 1/2 | -1800 | 3.29 | 0.037 | 115 | CAFM | 0.17 | [40] |
| | $CaPtO_3$ | $Pt^{4+}$: $5d^6$ ($t_{2g}^6$) | 0 | - | - | - | - | D | - | [43,44] |
| Pv | $CaRuO_3$ | $Ru^{4+}$: $4d^4$ ($t_{2g}^4$) | 1 | -68 | 2.2 | - | - | CWP | Metallic | [19] |
| | $CaRhO_3$ | $Rh^{4+}$: $4d^5$ ($t_{2g}^5$) | 1/2 | -650 | 3.52 | - | - | CWP | Metallic | [28,29] |
| | $CaIrO_3$ | $Ir^{4+}$: $5d^5$ ($t_{2g}^5$) | 1/2 | - | - | - | - | PP | Metallic | [40] |

[a] PPv: post-perovskite; Pv: perovskite, [b] $M_0$: spontaneous magnetization, [c] CWP: Curie-Weiss-like paramagnetism, CAFM: canted AFM, PP: Pauli paramagnetism, D: diamagnetism, [d] $E_a$: activation energy measured in electrical transport measurement. VRH: variable range hopping conduction.

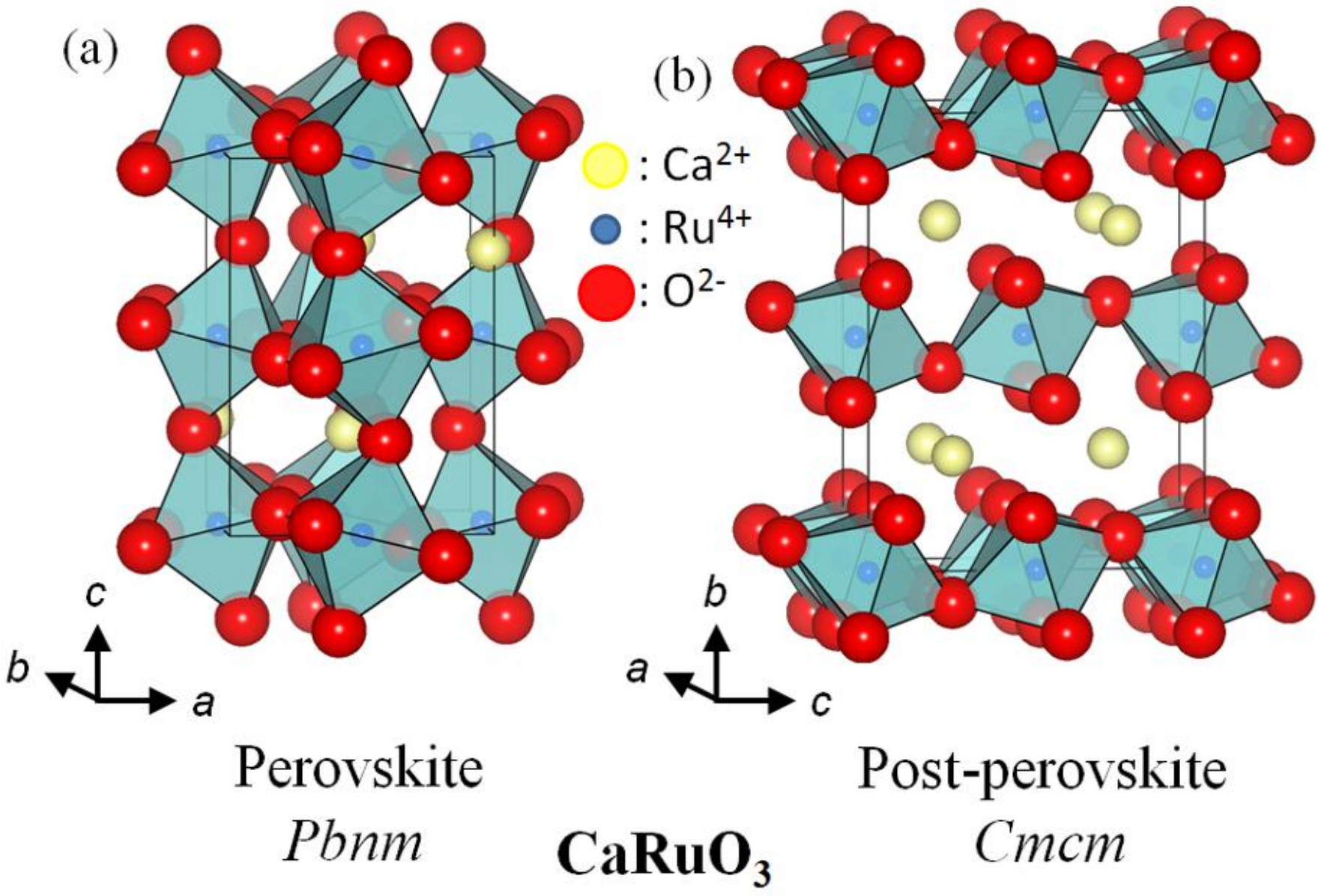

Fig. 1 Sketch of the crystal structures of (a) the perovskite and (b) the post-perovskite $CaRuO_3$ [1]. The figures used the same drawing scale. The post-perovskite phase is 1.7 % denser in volume than the perovskite phase [1]. The crystal structures are drawn by using VESTA [3].

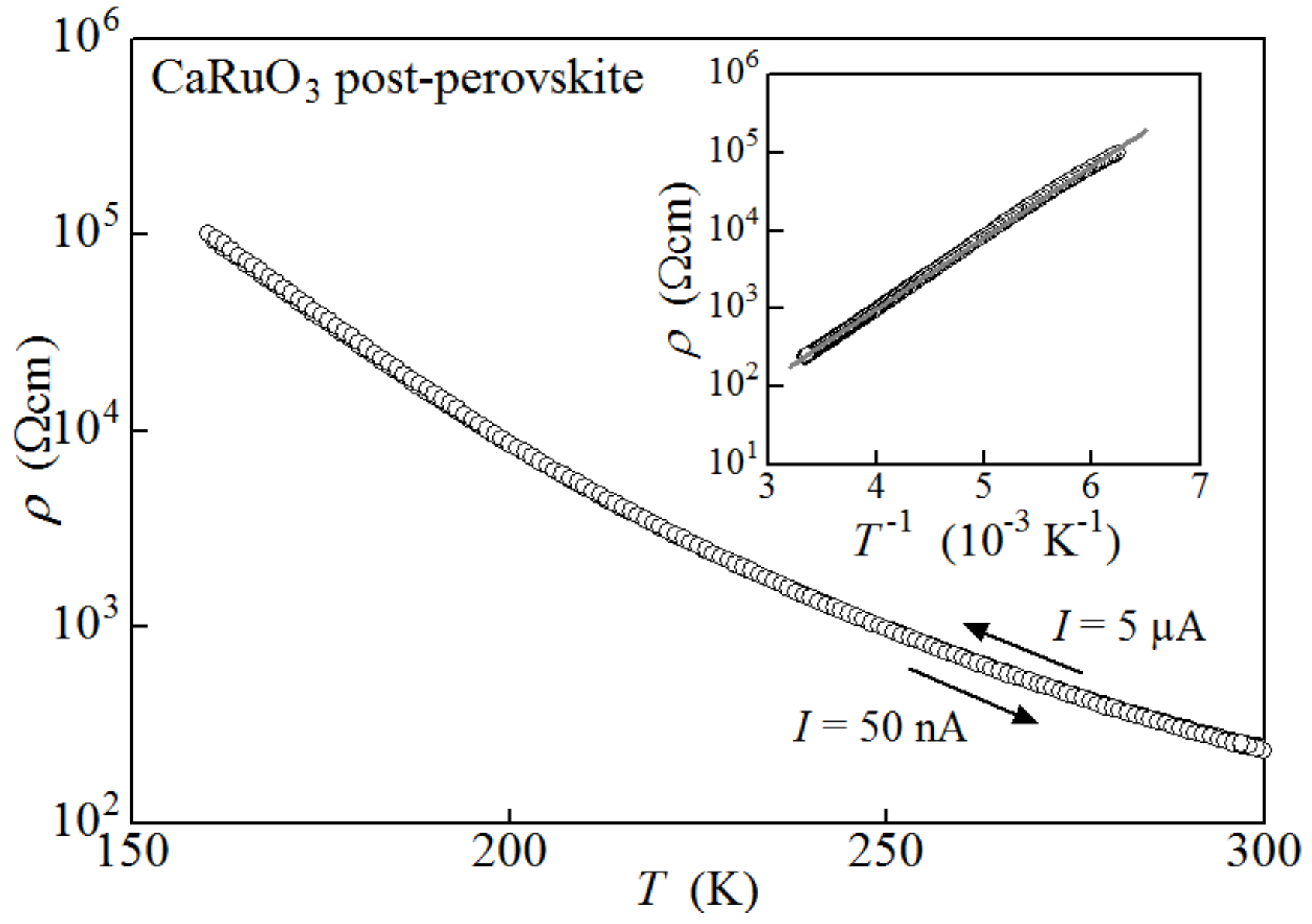

Fig. 2 $T$ vs. $\rho$ of a polycrystalline and dense ($d_{obs}/d_{cal}$ > 0.97) pellet of the post-perovskite $CaRuO_3$. Inset shows Arrhenius plot of the data.

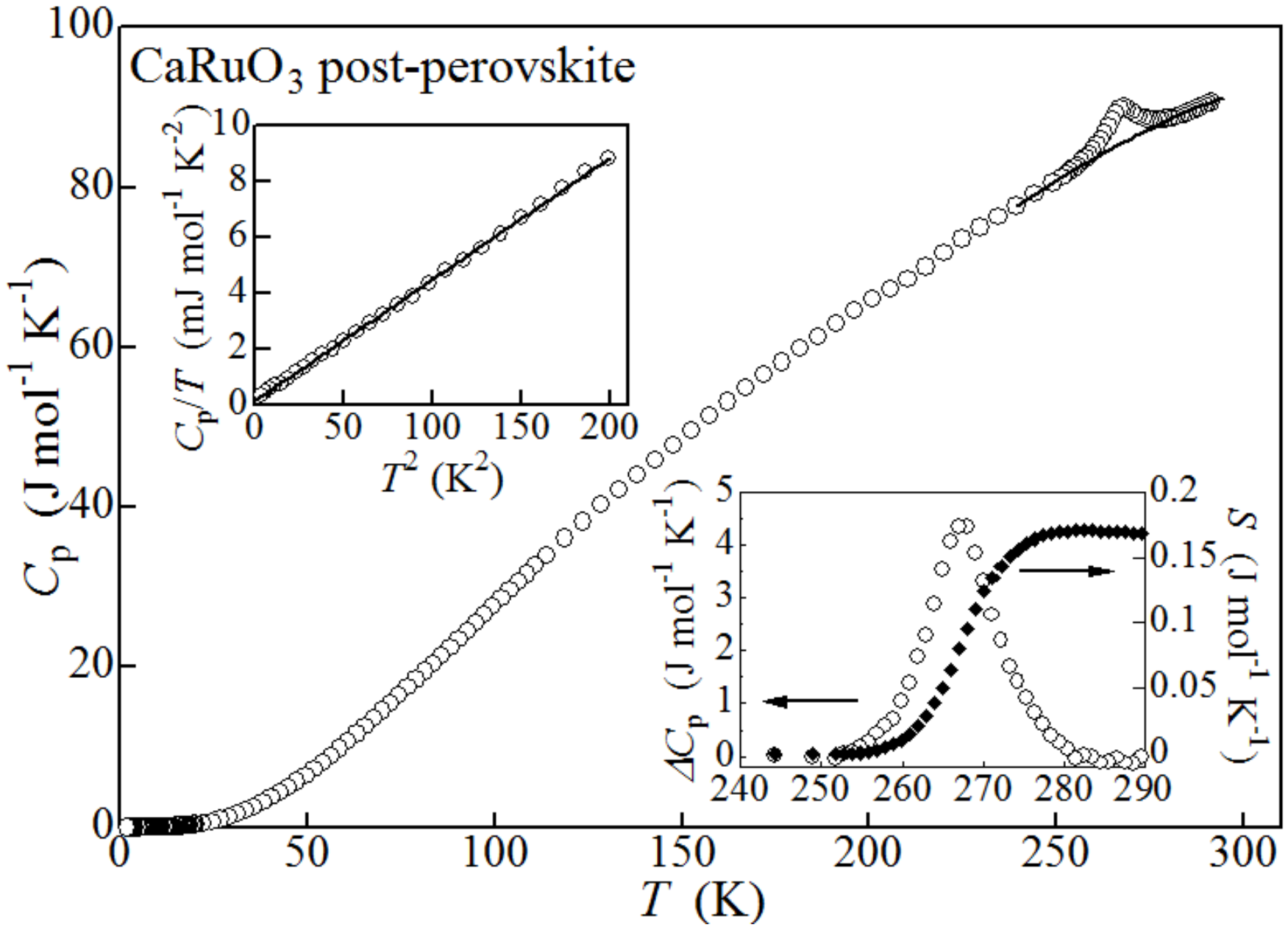

Fig. 3 $C_p$ vs. $T$ (main panel) and $C_p/T$ vs. $T^2$ (upper inset) of the post-perovskite $CaRuO_3$. Lower inset shows an estimation of the transition entropy at ~270 K.

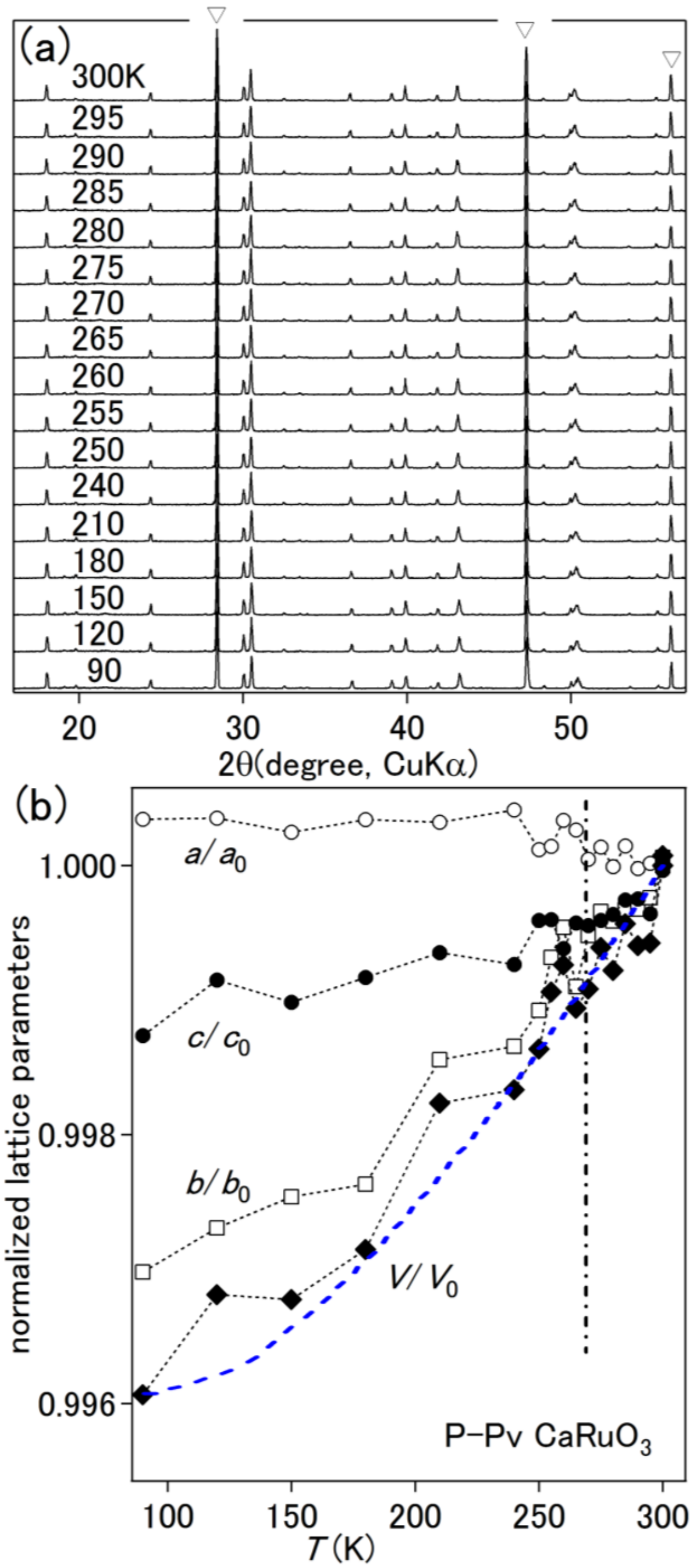

Fig. 4 (a) Low temperature XRD patterns and (b) temperature dependence of the normalized lattice parameters of the post-perovskite $CaRuO_3$ (*Cmcm*). The 300 K parameters $a_0$ = 3.115(1) Å, $b_0$ = 9.826(1) Å, $c_0$ = 7.297(1) Å, and $V_0$ = 223.3(1) Å$^3$ were used in the plot. The open triangles correspond to Si peaks. The broken line corresponds to the magnetic ordering temperature of the post-perovskite $CaRuO_3$ and the fat dotted curve indicates the $V/V_0$ curve for the perovskite $CaRuO_3$ (taken from Ref. 33).

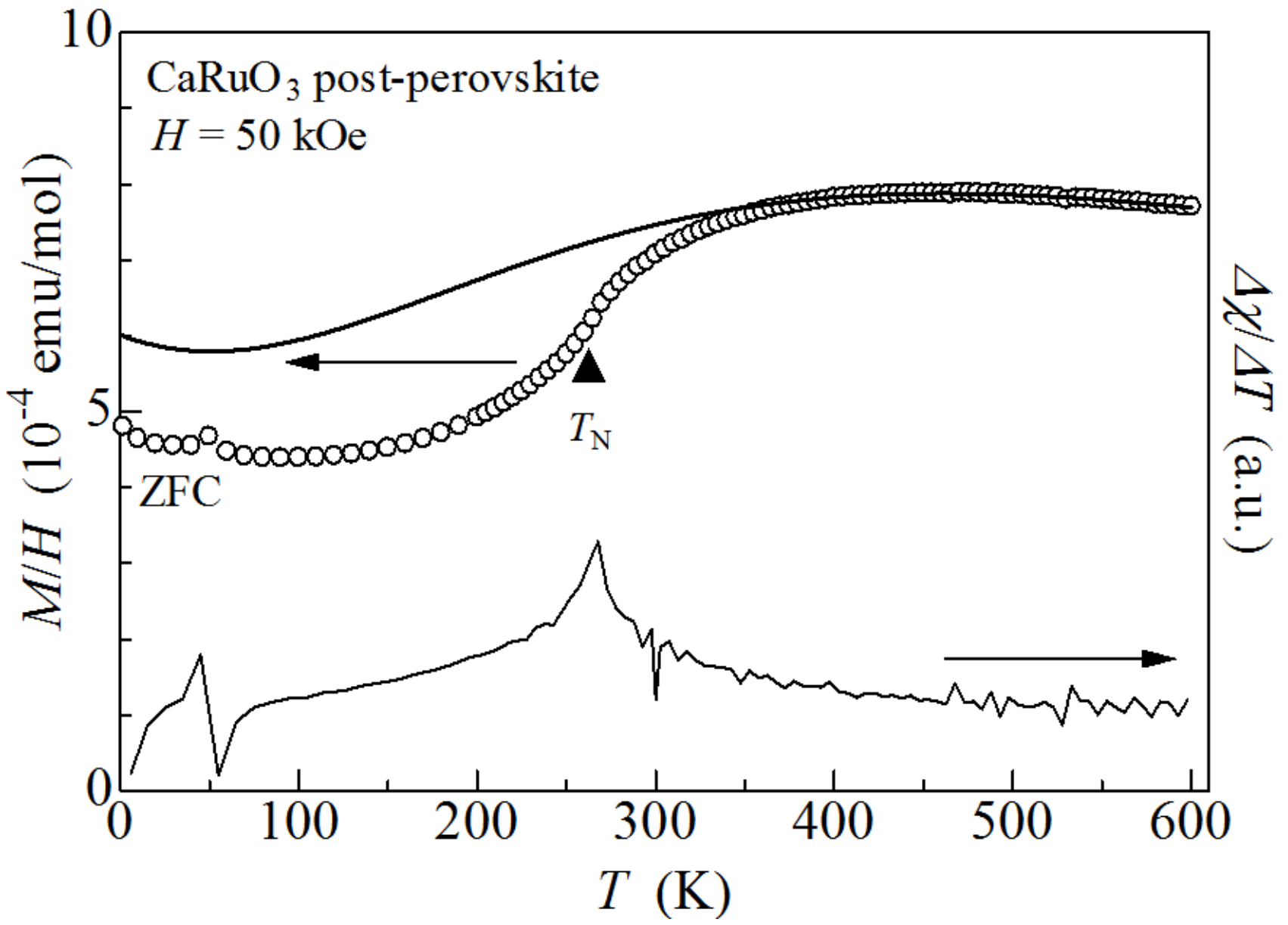

Fig. 5 $T$ dependence of the magnetic susceptibility of the post perovskite $CaRuO_3$ measured at 50 kOe and a theoretical curve (solid curve) fitting to the data. Differential curve at the bottom indicates an anomaly at ~270 K.

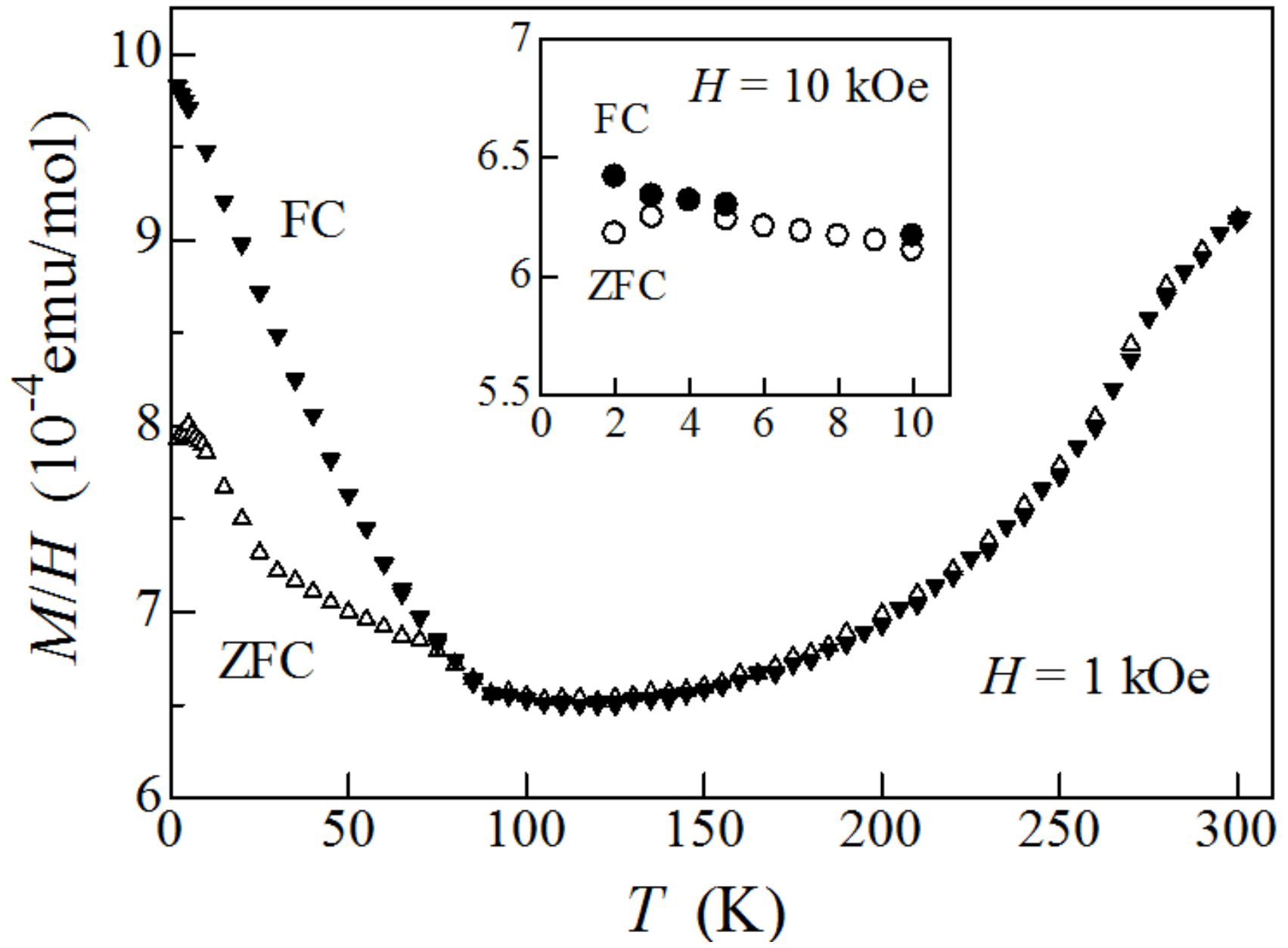

Fig. 6 Thermal hysteresis of the magnetic susceptibility of the post perovskite $CaRuO_3$ measured at 1 kOe (main panel) and 10 kOe (inset).

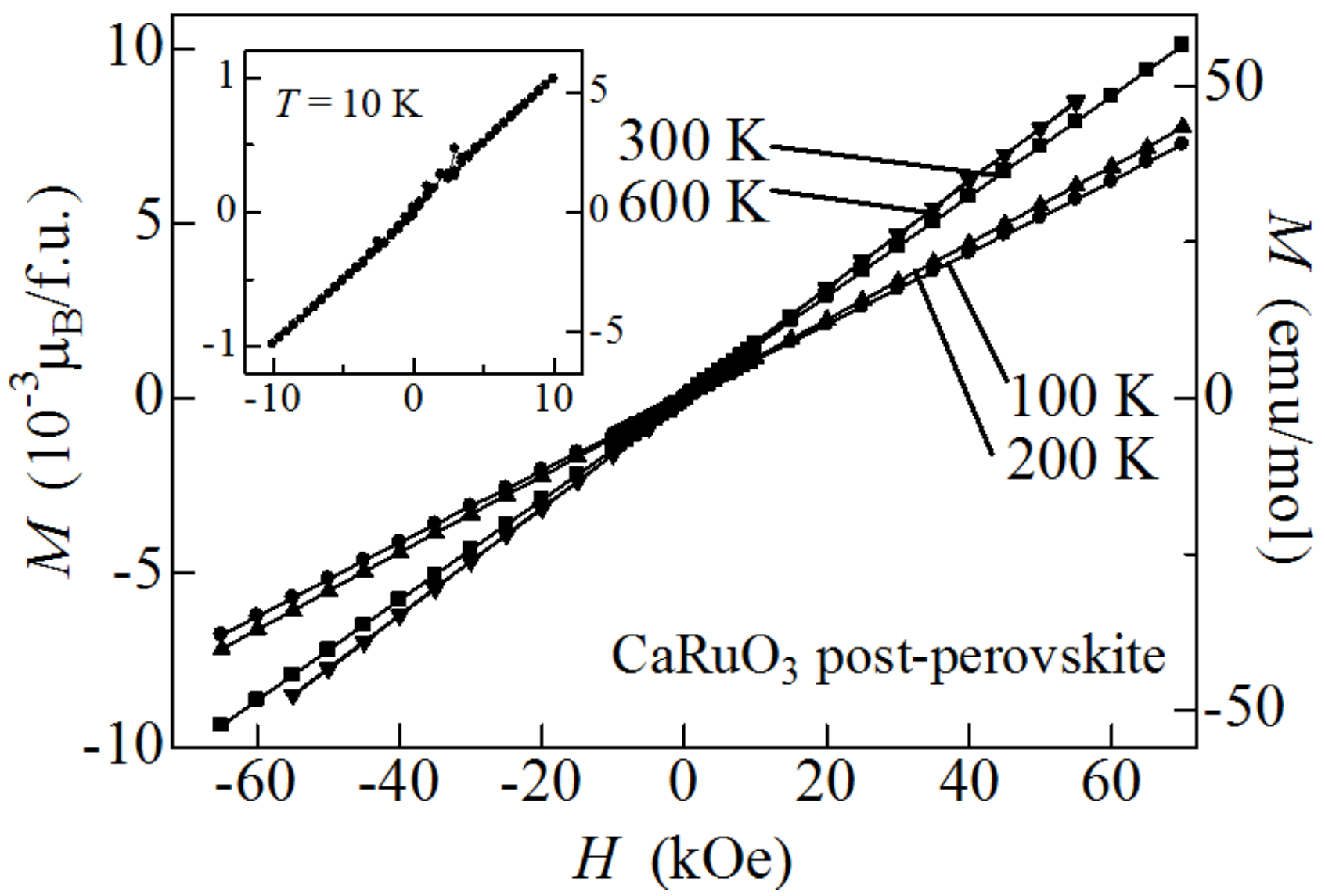

Fig. 7 Isothermal magnetization at various temperature of the polycrystalline post-perovskite $CaRuO_3$. Inset shows the data measured at 10 K.

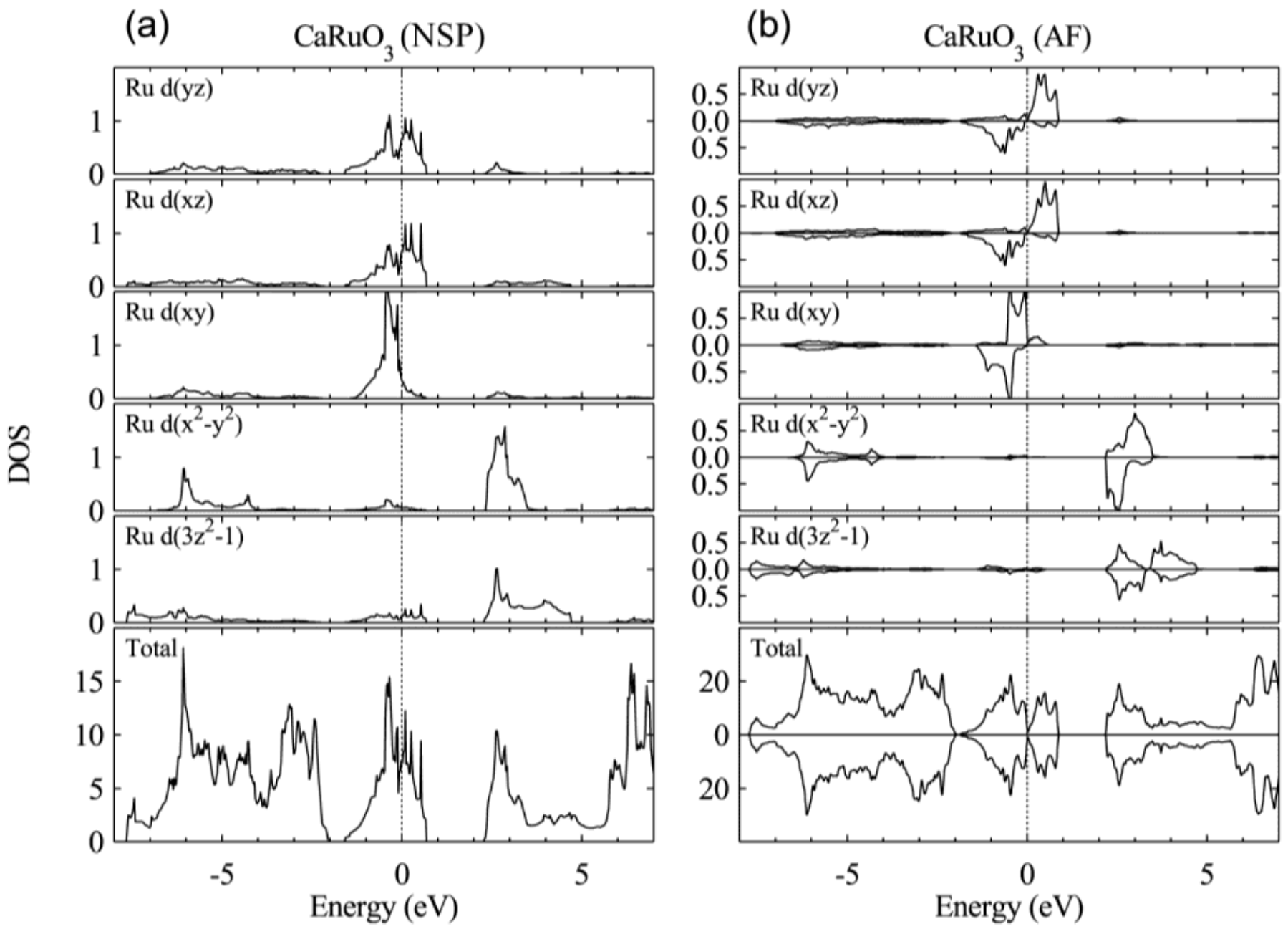

Fig. 8 Total and partial densities of states (DOS) for the post-perovskite $CaRuO_3$. We use a local coordinate system which has *z*-axis along Ru-O1 bond and *x*-axis approximately along Ru-O2 bond. (a) non spin-polarized states, (b) antiferromagnetic states.